\documentclass[
12pt, % Main document font size
a4paper, % Paper type, use 'letterpaper' for US Letter paper
oneside, % One page layout (no page indentation)
BCOR5mm, % Binding correction
]{scrartcl}
 \makeatletter
\DeclareOldFontCommand{\rm}{\normalfont\rmfamily}{\mathrm}
 \DeclareOldFontCommand{\sf}{\normalfont\sffamily}{\mathsf}
 \DeclareOldFontCommand{\tt}{\normalfont\ttfamily}{\mathtt}
 \DeclareOldFontCommand{\bf}{\normalfont\bfseries}{\mathbf}
 \DeclareOldFontCommand{\it}{\normalfont\itshape}{\mathit}
 \DeclareOldFontCommand{\sl}{\normalfont\slshape}{\@nomath\sl}
 \DeclareOldFontCommand{\sc}{\normalfont\scshape}{\@nomath\sc}
 \makeatother
\usepackage[english]{babel}
\usepackage[T1]{fontenc} % Use 8-bit encoding that has 256 glyphs
\usepackage{times}
\usepackage[utf8]{inputenc}% Required for including letters with accents
\usepackage{newunicodechar}
\newunicodechar{〈}{\langle}

\usepackage{graphicx} % Required for including images
\graphicspath{{Figures/}} % Set the default folder for images

\usepackage{enumitem} % Required for manipulating the whitespace between and within lists
\newlist{steps}{enumerate}{1}
\setlist[steps, 1]{label = Step \arabic*:}

\usepackage{lipsum} % Used for inserting dummy 'Lorem ipsum' text into the template

\usepackage{amsmath,amssymb,amsthm} % For including math equations, theorems, symbols, etc

\usepackage{varioref} % More descriptive referencing
\usepackage{booktabs} % For \toprule, \midrule and \bottomrule

\usepackage{siunitx} % Formats the units and values

\usepackage{pgfplotstable} % Generates table from .csv

\usepackage{etoolbox}

\usepackage{lipsum}

\usepackage{float}

\usepackage{amsmath}

\usepackage[T1]{fontenc}

\usepackage{natbib}

\usepackage{listings}

\usepackage{caption}

\usepackage{longtable}

\usepackage{filecontents}  

\usepackage{amsfonts}

\usepackage[colorinlistoftodos]{todonotes}

\usepackage{algorithm}

\usepackage{algpseudocode}

\usepackage{lineno}

\usepackage{etoolbox}

\usepackage{blindtext}

\usepackage{enumitem}

\usepackage{booktabs}

  \usepackage{url}

\usepackage[font={normal,it}]{caption}

\theoremstyle{definition} % Define theorem styles here based on the definition style (used for definitions and examples)

\theoremstyle{plain} % Define theorem styles here based on the plain style (used for theorems, lemmas, propositions)

\theoremstyle{remark} % Define theorem styles here based on the remark style (used for remarks and notes)

\usepackage{authblk}
\usepackage{chngcntr}
\counterwithin{figure}{section}
\PassOptionsToPackage{hyphens}{url}
\usepackage{breakurl}
\usepackage{breakcites}
\usepackage{caption}
\usepackage{listings}
\date{} % An optional date to appear under the author(s)

\let\clsCenter\Center\let\clsendCenter\endCenter
\let\Center\undefined\let\endCenter\undefined
\usepackage{ragged2e}
\let\Center\clsCenter
\let\endCenter\clsendCenter

\newcommand\SentenceCase[1]{%
  \caselower{}%
  \capitalize{\thestring}%
}

\title{A Grouping Genetic Algorithm for Joint Stratification and Sample Allocation Designs} 
\author
{Mervyn O'Luing,$^{1}$ Steven Prestwich,$^{1}$ S. Armagan Tarim$^{2}$
}
\usepackage{subfig}
\newsubfloat{figure}
\usepackage{bibentry}

\begin{document}

\maketitle
%----------------------------------------------------------------------------------------
%	ABSTRACT
%----------------------------------------------------------------------------------------
\begin{center}
    %\vspace{0.9cm}
    \textbf{Abstract}
\end{center}

 % This section will not appear in the table of contents due to the star (\section*)

Predicting the cheapest sample size for the optimal stratification in multivariate survey design is a problem in cases where the population frame is large. A solution exists that iteratively searches for the minimum sample size necessary to meet accuracy constraints in partitions of atomic strata created by the Cartesian product of auxiliary variables into larger strata.  The optimal stratification can be found by testing all possible partitions.  However the number of possible partitions grows exponentially with the number of initial strata.  There are alternative ways of modelling this problem, one of the most natural is using Genetic Algorithms (GA).  These evolutionary algorithms use recombination, mutation and selection to search for optimal solutions.  They often converge on optimal or near-optimal solution more quickly than exact methods.  We propose a new GA approach to this problem using grouping genetic operators instead of traditional operators.  The results show a significant improvement in solution quality for similar computational effort, corresponding to large monetary savings.


%----------------------------------------------------------------------------------------
%	AUTHOR AFFILIATIONS
%----------------------------------------------------------------------------------------

{\let\thefootnote\relax\footnotetext{\textsuperscript{1} \textit{Insight Centre for Data Analytics, Department of Computer Science, University College Cork, Ireland.} Email:{mervyn.oluing@insight-centre.org},{ steven.prestwich@insight-centre.org }}

{\let\thefootnote\relax\footnotetext{\textsuperscript{2} \textit{Department of Management, Cankaya University, Ankara, Turkey.} Email: {at@cankaya.edu.tr }}}

%----------------------------------------------------------------------------------------

%\newpage % Start the article content on the second page, remove this if you have a longer abstract that goes onto the second page

%\setcounter{tocdepth}{2} % Set the depth of the table of contents to show sections and subsections only

%\tableofcontents % Print the table of contents

%\listoffigures % Print the list of figures

%\listoftables % Print the list of tables

%----------------------------------------------------------------------------------------
%	INTRODUCTION
%----------------------------------------------------------------------------------------
\section{Introduction}

Survey objectives typically require a sample that is sufficiently representative of the relevant population characteristics. Samples taken from a stratification must meet accuracy constraints in order to satisfy these objectives. The problem of sample allocation for surveys with multiple variables was first discussed by \citep{neyman1934two}, whereas optimum stratification in survey sampling was first discussed by \citep{dalenius1950problem}. %\citep{ballin2013joint} state that: \emph{Stratified sampling is a widely adopted design that may ensure savings in costs and gains in accuracy of estimates, when stratification variables are available in the sampling frame.} In this paper variables are synonymous with variables. 
By variable we mean the set of values or measurements recorded for a given characteristic from a population unit in a survey. In {\it stratified random sampling\/} simple random samples are extracted from non-overlapping subpopulations or strata in a population \citep{cochran1977sampling}. It is used to achieve savings in costs while obtaining information for each stratum and preserving the accuracy of the target estimates in terms of variance. Our analysis in this paper is limited to partitioning a Cartesian product of the set of values in all auxiliary variables (in future we will just say the \emph{Cartesian product of auxiliary variables}) into larger groupings or strata and searching for the minimum sample size from these partitions necessary to meet precision constraints.

\citep{cochran1977sampling} outlines that stratification is useful where different levels of precision are required for respective subdivisions of a population. There is also an administrative purpose for the use of strata where reliable statistics are sought for various regions or provinces. Strata can also be used to apply relevant techniques to survey a population in accordance with certain characteristics. For example one interviewer might be adequate for a densely populated centre such as a hotel, whereas sparsely populated rural areas may require additional considerations such as more interviewers, over-time payments or travel costs. Similarly \citep{cochran1977sampling} notes that in sampling a business population it may be useful to put large firms into the same strata, separate to smaller firms which might require some form of area sampling. Another reason for stratification is where a heterogeneous population may be split into internally homogeneous strata, where the measurements vary little inside each stratum. Consequently, a small sample from a stratum can give a precise estimate of the stratum mean. The estimates of the mean from each stratum can be combined into a precise estimate of the mean for the whole population. 

All of the above reasons for stratification mean that once the heterogeneous strata are established they are physically tangible in the sense that measurements can be obtained for the population of interest to the survey. The sample allocation is optimised so that it can be expected that the estimates of samples gathered from each stratum will fall within the pre-set variance constraints. Published research in this area includes the work of  \citep{cochran1977sampling,bethel1985optimum,bethel1989sample,chromy1987design,huddleston1970optimal,kish1976optima, stokes2004using,day2006application,day2010multi,diaz2008multi,kozak2010stochastic,kozak2008stratified,khan2010optimal}. Bethel and Chromy proposed similar convex programming algorithms to solve the Neyman optimal allocation to strata where a number of variables are of interest. The Bethel algorithm uses the Kuhn Tucker theorem and Lagrangian multipliers and is proven to converge quickly on the optimum allocation, if it exists. The Chromy algorithm, which also uses the Kuhn Tucker theorem and Lagrangian multipliers or Cauchy-Schwarz solutions, is simpler than the Bethel algorithm and has evidence of good convergence but this has not been proven. However, its notation can be adapted to fit-within the operation of the Bethel algorithm and the two have been combined into the \emph{bethel.r} function (also known as the Bethel-Chromy function)\citep{barcaroli2014samplingstrata}, which is discussed further below. It is used to estimate minimum sample size for the variance constraints for a fixed stratification, whether they are physically tangible like those outlined above, or constructed like those discussed further down.  

Research has also been conducted on two-stage sampling which optimises strata initially, and subsequently addresses sample allocation in the survey design. The initial step focuses on determining methods of constructing strata using auxiliary variables which are correlated to the target variables to establish expected precision levels. Work in this area includes \citep{dalenius1959minimum}, \citep{singh1971approximately}, \citep{hidiroglou1986construction}, \citep{allee1988stratification}, \citep{gunning2004new}, \citep{khan2008determining},\citep{CIS-196217},\citep{lednicki2003optimal,briggs2000strat2d,kozakapplication,lu2002multi,reddy2016procedure,rizvi2000approximately,schneeberger1985optimum,thomsen1977effect}.  

Finally there is research into the area of joint stratification and optimising sample size. Strata for both the univariate and the multivariate cases are constructed using correlated auxiliary variables. The process involves jointly considering minimum sample size for constructed strata in such a way as to provide the stratification that gives the minimum sample size required to meet constraints.  It is possible to sub-divide the population into domains, e.g. areas or regions before estimating minimum sample size for constructed strata in each domain. Techniques used to solve the joint sample allocation and stratification problem include evolutionary and tree-based approaches of selecting the best solution. Work in this area includes \citep{kozak2007modern}, \citep{keskinturk2007genetic} , \citep{baillargeon2009general}, \citep{baillargeon2011construction} who address the univariate scenario and then there is \citep{benedetti2005tree} and \citep{ballin2013joint} who deal with multivariate stratification scenarios.

%The total sample size estimated from the un-partitioned set of atomic strata could be very large, especially if the sampling frame is large and number of atomic strata is high.   
\citep{ballin2013joint} use a Genetic Algorithm (GA) which is a modified version of \emph{rbga.r} in the \emph{genalg} R package to explore the search space of partitions of constructed strata to find the minimum sample size. Each partition contains subsets of constructed strata (henceforth called \emph{atomic strata}) and these subsets represent stratifications of the atomic strata (henceforth called \emph{strata}). Each partition is a candidate solution. GAs often converge quickly to optimal or near optimal solutions, and are particularly good at navigating rugged search spaces containing many local minima.  The work of \citep{ballin2013joint} has been applied to the following surveys carried out by the Italian National Institute of Statistics (ISTAT):

\begin{itemize}

\item 2003 Italian Farm Structure Survey 

\item 2010 Monthly milk and milk products survey

\item Economic outcomes of agricultural holdings survey

\item Structure and production of main wooden cultivations survey

\item Survey on forecasting of some herbal crops sowing

\end{itemize}

This paper presents research aimed at further developing their work by exploring an alternative GA with genetic operators that we claim are better suited to this application.  We support our claim by comparing computational results on publicly-available test data. The genetic algorithm in the \emph{SamplingStrata} 1.2 package \citep{barcaroli2018optimization} has been updated following the analysis we provide in this paper. %Our purpose merely is to explore an alternative genetic algorithm for solving the problem to the genetic algorithm laid out in \citep{ballin2013joint} and \citep{barcaroli2014samplingstrata}. We assume a knowledge of the method of stratification from partitions of the Cartesian product of categorical auxiliary variables as well as a practical familiarity with the R package \citep{barcaroli2014samplingstrata}.

%We are taking a further step into a branch of data analytics in which computer science techniques are used to solve common statistical problems. There will be some cross-over in terminology. For example in describing the stratification problem we intend that population refers to the (potentially infinite) set of stratified units from which a sample is drawn in order to make inferences about that population. However, in describing the genetic algorithms we intend that population refers to a set of candidate solutions which are iteratively evolved through a process of fitness testing, ranking, recombination and mutation. 
 
%----------------------------------------------------------------------------------------
%	Problem Description
%----------------------------------------------------------------------------------------

\section{Problem Description}

We refer to the problem description laid out in \citep{ballin2013joint} and \citep{barcaroli2014samplingstrata}. This problem requires a population dataset in which there are $N$ records. For each of these records there are target variables (also known as target features), i.e., those which are of interest to the survey. These are described as follows:

\begin{equation}
\begin{array}{rrclcl}

 Y_g\ (g=1,\ldots,G)

\end{array}
\end{equation}

where $Y$ is a target variable and $G$ is the total number of target variables. 

A subset of the dataset containing the most correlated auxiliary variables is also selected. %These can be selected by various methods which we will not go into in this paper, as our focus is on stratification after both the target and auxiliary variables are selected, however the selection process can include expert knowledge of the data, stepwise multivariate regression, principal component analysis, or a correlation matrix. The selection of correlated variables is particularly useful in datasets where there are a large number of auxiliary variables as it has a practical implication on calculation of the Cartesian product, and the running of the algorithm as it is outlined below. 
%The auxiliary variables can be nominal, ordinal or continuous. Continuous variables will be converted to categorical variables using the k-means clustering technique. 
These auxiliary variables are formally described as follows:

\begin{equation}
\begin{array}{rrclcl}

 X_m\ (m=1,\ldots,M)

\end{array}
\end{equation}

where $X$ is an auxiliary variable and $M$ is the total number of auxiliary variables.  

The Cartesian product of the auxiliary variables creates atomic strata for each domain (e.g. region, district, gender, social group, etc.). 

\begin{equation}
\begin{array}{rrclcl}

CP = X_1 \;  \times \;  X_2\;   \times \ldots\; \times \;X_M

\end{array}
\end{equation}

Atomic strata which are effectively empty sets (no corresponding values in the $Y_G$ target variables) can be removed from the total. The number of possible atomic strata is described as follows:

\begin{equation}
\begin{array}{rrclcl}

K = \prod_{m=1}^M k_m - I^* 

\end{array}
\end{equation}

where $K$ is the number of atomic strata, and $I^*$ is the number of impossible or absent number of combinations of values created by the Cartesian product of the $M$ auxiliary variables.   $k_m$ refers to an atomic stratum $k$ in auxiliary variable $m$.

The Cartesian product results in the set of $L$ atomic strata of size $K$.
\begin{equation}
\begin{array}{rrclcl}

L =\left\{l_1,l_2,\ldots,l_K\right\}

\end{array}
\end{equation}

After the creation of atomic strata the next step is to consider all possible partitions of the atomic strata and for each of these partitions:  estimate the minimum sample size necessary to meet accuracy constraints. We then search for the partition that enables the global minimum sample size for the precision constraints. %The goal is that this sample size will be lower than that achievable from stratification by domain, single or two-stage clustering, a combination of stratification by domain and clustering, or systematic sampling by domain - for the same precision constraints.
\\
%This example uses a bivariate scenario, but the process is also applicable to a multivariate scenario.  The Cartesian product of the auxiliary variables in a multivariate scenario is:

%An example of atomic strata resulting from a cartesian product in a mutivariate scenario is included below. The particular experiment using the iris dataset from which the table is prepared will be discussed in detail in section 6, however the table is included here because we will refer to the eight atomic strata when we discuss our grouping genetic algorithm in section 4. Most datasets will be much larger than the iris dataset (which has only 150 records) and the maximum number of elements that can be created using the Cartesian product of auxiliary variables is potentially enormous. 

%As described above, each common element created by this cross-product is grouped together into an atomic stratum.  
%There may not be a corresponding value in the target variables for each atomic stratum. Thus the a

%The atomic strata are then labelled as follows:

%\begin{equation}
%\begin{array}{rrclcl}
%
%L =\left\{l_1,l_2,\ldots,l_K\right\}
%
%\end{array}
%\end{equation}
%
%where $L$ is the set of atomic strata.  

The set of all possible partitions $P$ of the $L$ atomic strata is defined by $\left\{P_1 , P_2,\ldots,P_B\right\}$. 

Each partition is a candidate solution. The number of possible partitions is known as the search space, because, if it was feasible in terms of time and cost, the full list of candidate solutions would be evaluated for the optimum solution. However, given that the set of partitions grows exponentially with the set of atomic strata, search techniques such as genetic algorithms are used as they can find an optimum solution without evaluating all possible solutions. There is more information on the genetic algorithm we use later in the paper. \citep{ballin2013joint} provide a detailed report of the partitioning problem and describes how to use the Bethel-Chromy algorithm to select the best partition for a given domain. We base our outline of the problem on the brief description provided by \citep{barcaroli2014samplingstrata}.

The set of atomic strata $L$ is partitioned into subsets of strata called $H$. Please note the distinction between the set of $L$ atomic strata (of length or size $K$) and strata $H$ (or set of subsets of $L$) which results from a given partition of the set $L$. Assuming $H$ strata, $N_h$ is the population size for each different $Y_g$ target variable in each stratum $h$. Similarly $S_{h,g}^2$ are the variances of the $G$ different target variables $Y$ in each stratum $h$, where $ h = 1,\ldots,H$ and $ g = 1,\ldots,G$. We assume a simple random sampling of $n_h$ units without replacement in each stratum.

The estimator of the population total $\left(\hat{T}\right)$ is:

\begin{equation}
{E\left(\hat{T}_g\right)}=\sum_{h=1}^{H}N_h\bar{Y}_{h,g}; (g=1,\ldots,G)
\end{equation}
\\
where:\\

$\bar{Y}_{h,g}$ is the mean of the $G$ different target variables $Y$ in each stratum $h$
\\ 

The variance of the estimator of the total $\left(\hat{T}\right)$ is given by:

\begin{equation}
\begin{array}{rrclcl}

\mbox{VAR}\left(\hat{T}_g\right)=\sum_{h=1}^{H}N_h^2\left(1-\frac{n_h}{N_h}\right) 

\frac{S_{h,g}^2}{n_h} \;\;\; (g=1,\ldots,G)

\end{array}
\end{equation}

The upper limit of expected variance or precision $U_g$ is expressed as a coefficient of variation $CV$ for each $\hat{T}_g$, which normalises the variability in measurements in the different target variables and means that the variance can be compared relatively:

\begin{equation}
\begin{array}{rrclcl}

CV\left(\hat{T}_g\right)= \frac{\sqrt{\mbox{VAR}\left(\hat{T}_g\right)}}{E\left(\hat{T}_g\right)}\leq U_g

\end{array}
\end{equation}

where $U_g$ is the expected sampling variance or precision for target variable $g$ expressed as a coefficient of variation.\\

The cost function can be defined as:

\begin{equation}
C\left(n_1,\ldots,n_H\right)=C_0 + \sum_{h=1}^{H}C_hn_h
\end{equation}
\noindent
where $C_0$ is the fixed cost and $C_h$ is the average cost of interviewing one unit in stratum $h$ and $n_h$ is the number of units, or sample, allocated to stratum $h$.\\ 
\noindent
As mentioned above, the set of $H$ strata will vary for each partition of the $L$ atomic strata. However, if the average cost of sampling a unit in the population is fixed this means that even though the set of $H$ and size of each $h$ might vary, the average sampling cost for a unit in $h$ remains fixed. Alternatively, the sampling costs for each atomic stratum might also be estimated after the set of atomic strata has been prepared. \citep{ballin2013joint} indicates that the expected cost of observing a unit in a given aggregate stratum can be estimated by averaging the costs in each contributing atomic stratum, weighted by their population. In our analysis we take the view that $C_0$ and $C_h$ are constant in the linear cost model therefore both can be utilised after the minimum sample size is found. Therefore, $C_0$  is set to $0$, and $C_h$ is set to $1$.\\

The problem is consequently summarised as follow:

\begin{equation}
\begin{array}{rrclcl}
\begin{cases}\min \rightarrow n = \sum_{h=1}^{H}n_h
\\
\\

CV\left(\hat{T}_g\right)\leq U_g

\end{cases}
\end{array}
\end{equation}

The problem is then solved using the genetic algorithm in \emph{SamplingStrata} to evolve a set of potential solutions which are evaluated to estimate the minimum sample size necessary to meet the precision constraints using the \emph{bethel.r} function and each of these solutions is ranked from lowest sample size to highest.

%------------------------------------------------

\section{Genetic Algorithms}

Genetic Algorithms (GAs) are a nature-inspired class of optimisation algorithms, modelled on the ability of organisms to solve the complex problem of adaptation to life on Earth. The variables of an optimisation problem are called {\it genes\/} and their values {\it alleles\/}. A candidate solution is a list of alleles called a {\it chromosome\/} and the objective function, which is maximised by convention, is called the chromosome's {\it fitness\/}. The search for fit chromosomes (solutions with high objective) uses two {\it genetic operators\/}: small random changes called {\it mutation\/}, equivalent to small local moves in a hill-climbing algorithm; and large changes called {\it crossover\/} in which the genes of two {\it parent chromosomes\/} are {\it recombined\/}. GAs often give more robust results than search algorithms based on hill-climbing, because of their use of recombination, and they have found many applications since their introduction in 1975 by John Holland.

The original genetic algorithm (GA) used for this problem, in the R Core
\cite{R} package \textit{SamplingStrata}
\citep{barcaroli2014samplingstrata}, is an elitist generational GA in
which the atomic strata $L$ are considered to be elements of a set (or
genes) for a standard one-point crossover strategy.  In each iteration
the best solutions are carried over to the next generation of
candidate solutions.  For each remaining candidate solution (or
chromosome) in the new generation two of the solutions from the
original are randomly selected.  These parents are recombined to form
one offspring and are mutated at a specified or default probability.
Classical genetic operators are used: point mutation and single-point
crossover.

However, the problem is an example of a {\it grouping problem\/}
(related to {\it packing\/}, {\it cutting\/} and {\it partitioning\/}
problems).  In such problems a set of objects must be partitioned into
disjoint groups.  The obvious problem representation is that used in
\textit{SamplingStrata}: one gene per object, with an allele (value) for each
group, and standard genetic operators can be applied to this
representation (as was done in \textit{SamplingStrata}).

Unfortunately, standard genetic operators are known to perform poorly
on grouping problems.  The reason is that the representation contains
a great deal of {\it symmetry\/}: given a grouping, permuting the
group labels yields an equivalent grouping.  Symmetry has a damaging
effect on GAs because recombining similar parent groupings might yield
a very different offspring grouping, violating the basic GA principle
that parents should tend to produce offspring with similar fitness.
This effect can be alleviated by designing more complex genetic
operators and problem models \cite{GalHao} or by clustering techniques
\cite{PelGol}.  We shall follow the former approach by designing a
special GA called a {\it grouping genetic algorithm\/} (GGA)
\citep{falkenauer1998genetic}.  GGAs have been shown to perform far
better than standard GAs on grouping problems, so it might be a better
approach for our problem than that of \citep{ballin2013joint}.

\section{Grouping Genetic Algorithms}

In a GGA a chromosome represents a list of groups, for example
\[
(\{1\},\{2\},\{3,6\},\{4\},\{5\})
\]
Here the numbers are objects and the chromosome contains 5 groups
(indicated by $\{\ldots\}$).  Recombination works as follows.  Suppose
we have two parent chromosomes:
\[
(\{1\},\{2\},\{3,6\},\{4\},\{5\}) \hspace{10mm}
(\{1,2\},\{3\},\{5,6\},\{4\})
\]
To generate an offspring, choose a section of each of the two parents,
for example $(\{2\},\{3,6\})$ from the first chromosome and
$(\{5,6\},\{4\})$ from the second one.  Then inject the section from
the first chromosome into the second chromosome, at the start of its
section: for example inject $(\{2\},\{3,6\})$ into the second
chromosome just before group $\{5,6\}$, giving an offspring
\[
(\{1,2\},\{3\},\{2\},\{3,6\},\{5,6\},\{4\})
\]
This is not a solution because some items occur twice, so remove
repeated items that were already in the chromosome before injection:
\[
(\{1\},\{\},\{2\},\{3,6\},\{5\},\{4\})
\]
Finally, delete any empty groups:
\[
(\{1\},\{2\},\{3,6\},\{5\},\{4\})
\]
To generate the other offspring (if the chosen GA requires 2
offspring) inject the section $(\{5,6\},\{4\})$ from the second
chromosome into the first one just before its section:
\[
(\{1\},\{5,6\},\{4\},\{2\},\{3,6\},\{4\},\{5\})
\]
Again delete repeated items that were there before injections:
\[
(\{1\},\{5,6\},\{4\},\{2\},\{3\},\{\},\{\})
\]
and delete empty groups:
\[
(\{1\},\{5,6\},\{4\},\{2\},\{3\})
\]
Mutation is simple: items can be moved between groups, or moved to new
groups (in some applications this must respect {\it capacity
  constraints\/}).  GGAs typically have an additional operator called
{\it inversion\/}.  This selects a section of the chromosome and
reverses it.  For example
\[
(\{1\},\{2\},\{3,6\},\{4\},\{5\})
\]
might become
\[
(\{1\},\{2\},\{5\},\{4\},\{3,6\})
\]
This does not affect the packing represented by the chromosome, but it
changes the future effects of crossover.  The idea is that some
orderings might be better than others because they are more likely to
keep together groups that belong together in good packings.

\section{GGA for our problem}

There is no single canonical GGA and they tend to be tailored for
specific applications, but most have the above features in common.  We
now design a tailored GGA for our problem, adapted from one designed
for the {\it Blockmodel Problem\/} \citep{james2010grouping}.  We fit
it into a modified version of the function called \emph{rbga.r} from
the \textit{genalg} R package \citep{willighagen2005genalg}.  It is
designed to work with the other functions in \textit{SamplingStrata},
and is applied to the joint stratification and optimum sample size
problem as follows.

Firstly we generate a random initial population.  We obtain the full
set of viable atomic strata of size $K$ from the Cartesian product of
the auxiliary variables (see above).  A population, or set of candidate
solutions, of a predetermined size $p$, is randomly generated for the
$K$ atomic strata. Each candidate solution (or individual) is a vector representing a stratification of the atomic strata %{\bf
 % [Steve: ``stratification of strata'' seems redundant but I can't see
 %   how to reword it]}.  
The stratification in the vector is generated
by $K$ integers randomly generated from the interval $[1,K]$, so the
genome of each individual contains integers ranging in value $v_k$
from $1<v_k<K$.  The frequency with which the integers occur is used
to define the groups for the atomic strata. % {\bf [Steve: I'm not sure
    %I understand this representation.]} 
%For example let's consider the 8 atomic strata constructed from the iris dataset example. If we select a random integer between 1 and 8 for each atomic strata (i.e. 8 times) we might get a vector like this (4,1,1, 4, 8, 7, 8,1). We repeat this for each chromosome (individual or candidate solution) in the population (set of candidate solutions). This corresponds to a stratification $P(v)$ of the atomic strata in each individual in the
%population.  The vector $v$, has $K$ elements,$v=[1,...,K]$,
%where each element is an atomic stratum earlier defined as $l_k$ ,is
%now grouped in accordance with the integer value. We will further illustrate this in section 5.1.

The fitness for each chromosome in the population is calculated by
estimating the minimum sample size required to meet constraints.  The
resulting sample size or cost is then ranked from lowest to highest
and the population is sorted accordingly.  $C_0$ is the fixed cost and
is excluded from our cost function as it is unrelated to sample-size.
$C_h$ is the average sampling unit cost.  For the purposes of our
analysis below we assume that because $C_h$ is constant in the linear
cost model it is not essential to the selection of sample size.  If
$C_h$ is set to $1$, then the cost function will simply calculate the
minimum sample size required to meet the constraints.  $C_0$ is set to
$0$, and $C_h$ is set to $1$ in the experiments described below.
$n_1,...,n_{H_P(v) }$ are the sample sizes for each stratum in the
partitioning of individual $P(v)$.

%Set the evaluation function to calculate the total cost or minimum
%sample size required to meet the constraints on the $G$ different
%target variables or estimates $\hat{Y}_g$ for each chromosome in the
%population.  Please refer to \citep{ballin2013joint} for a description
%of how the evaluation function works.  {\bf [Steve: I can't see how to
%    fit this into the above.]}

The GGA now evolves the population as follows.  It selects two parents
to generate an offspring.  This needs to be done for each remaining
individual in the population after the elite individuals have been
added, i.e.  $N_p-E$ times.  Create $N_p-E$ child chromosomes, by
randomly selecting two crossover points of the groups list in parent
$1$ and inserting the corresponding values from ‘parent 1’ into a copy
of ‘parent 2’, to create the child.  Remove any empty groupings from
each chromosome.  Empty groups can occur when the new values are
inserted into the child.  Rename the groups so that there is a
consecutive order to the group names: as mentioned above these are
integers, so it is useful to keep the group names as they were
originally.  Apply mutation to each value of the child, at a rate
which is determined by a preset mutation probability.

Evaluate the fitness of each individual in the new population and rank
them.  Calculate the number of individuals $E$ to be kept from one
generation to the next.  This is obtained by multiplying the elitism
rate $e$ by the population size, $p$ and rounding down to the nearest
integer: $E = \lfloor e \times p \rfloor$.  Then take the first $E$
individuals from the sorted population, which equates to the $E$ best
individuals carried forward to the next generation.

The fitness for each chromosome in the population is evaluated by
estimating the minimum sample size required to meet constraints.  The
population is then ranked in ascending order on the basis of which
chromosome enables estimation of the smallest sample size.  It selects
the $E$ best individuals on the basis of fitness value for carrying in
to the new population to be generated.  With a preset probability it
inverts each individual in the population.  Inversion is described in
more detail below.

Repeat until the last iteration, the output from the last iteration is
the population which gives the lowest sample size that has been found
from the search space of possible solutions at this stage.

Set the number of iterations which we wish to run the algorithm for.
This should be enough to give the GGA a chance to converge on the
optimum solution after the mutation and inversion likelihoods have
been applied.  If, however, the optimum solution is known beforehand
the algorithm can be set to stop at this point.

The mutation function is as described by \citep{cortez2014modern}.
For each atomic stratum of the child chromosomes a random number is
generated between 0 and 1.  If this number is less than the mutation
probability, the value or group number of the atomic stratum is
changed (by generating a new value from between 1 and the maximum
grouping number), otherwise it is not changed.

Recall that strata should be mutually disjoint but internally homogeneous subsets of the population. There are two levels of strata in this problem. The basic level is the atomic strata created from the cartesian product of auxiliary variables (see section 2). We intend to create higher level strata/subsets or groupings of atomic strata.  If each atomic stratum contains values that are the same or close in value then smaller sample sizes are needed for a precise estimate of the mean or total for the target variables. It follows also that grouping homogeneous atomic strata into strata will also require a smaller sample size to meet precision constraints. This is what leads to a good candidate solution, i.e. the strata are internally homogeneous but mutually disjoint and smaller samples sizes result. The GGA should more efficiently exploit good solutions in searching for an optimal solution.

\section{Comparing the Genetic Algorithms}

When our GGA was being developed its performance was compared to that of the original GA using experiments on the \emph{swissmunicipalities} data example provided in \citep{barcaroli2014samplingstrata}. We will report these findings in section 6. We used the same experiments so that we would have an idea of what results to expect and to see whether the GGA was showing an improved result for the same starting parameters. The \emph{swissmunicipalities} dataset represents a complete population that provides information about Swiss municipalities in 2003. Each municipality belongs to one of seven regions  which are at the NUTS-2 level, i.e. equivalent to provinces. Each region contains a number of cantons, which are administrative subdivisions. There are 26 cantons in Switzerland. The data, which was sourced from the Swiss Federal Statistical Office and is included in the \emph{sampling} and \emph{SamplingStrata} packages contains 2,896 observations (each observation refers to a Swiss municipality in 2003). They comprise 22 variables, details of which can be examined in \citep{barcaroli2014samplingstrata}.

The target estimates are the totals of the population by age class in each Swiss region. In this case, the $G$ target variables will be: 
\begin{itemize}[label={}]
\item Y1: number of men and women aged between 0 and 19, 
\item Y2: number of men and women aged between 20 and 39, 
\item Y3: number of men and women aged between 40 and 64, 
\item Y4: number of men and women aged 65 and over.
\end{itemize}
Given that 4 target variables have been selected there are only 18 remaining variables to choose the auxiliary variables from. However, it is useful to draw a selection of the most correlated variables from the 18, in such a way that they will be of interest to survey designers and this in turn ensures that the sample search space which arises from their cross-product will not be any bigger that what’s required. The $M$ auxiliary variables were chosen because they relate in some way to the use of land area. As they are continuous variables, they were converted into the categorical variables below, using a k- means univariate clustering method \citep{hartigan1979algorithm}:
\begin{itemize}[label={}]
\item X1: Classes of total population in the municipality. 18 categories.
\item X2: Classes of wood area in the municipality. 3 categories.
\item X3: Classes of area under cultivation in the municipality. 3 categories.
\item X4: Classes of mountain pasture area in the municipality. 3 categories.
\item X5: Classes of area with buildings in the municipality. 3 categories.
\item X6: Classes of industrial area in the municipality. 3 categories.
\end{itemize}
There are 7 regions, which we can use as administrative strata. They can be called domains which distinguishes them from the strata which both algorithms construct. The decision to stratify by region is because, as outlined above, we wished to replicate the experiment outlined in \citep{barcaroli2014samplingstrata}. However researchers can also compare the performance of our GGA against the \emph{SamplingStrata} GA on stratification by state for the American Community Survey PUMS dataset in section 7 below.

The selected target, auxiliary variables, domains as well as the municipal names variable are combined into a data frame. The municipal names variable is an identifier and will not be used to make the Cartesian product or calculate the minimum sample size. This data frame has the same number of observations as \emph{swissmunicipalities}, but fewer variables.

Atomic strata are then calculated from the Cartesian product of the auxiliary variables. Each stratum has a particular identification which is a concatenation of the common values arising from each valid cross product of the auxiliary variables. The expected number of atomic strata arising from the Cartesian product is 4,374, however not all combinations of values can be matched with corresponding values in the auxiliary variables, and as a result 641 valid atomic strata are selected. Recall that $C_0$ is set to 0, and $C_h$  is set to 1, which means that our cost function is 
\begin{equation}
C=n=\sum_{h=1}^HC_h n_h 
\end{equation}

where $ h = 1,\ldots,H$ and each $h$ is a subset of the set of atomic strata $L$

Our precision constraints are set as 0.05 for each target variable and for each domain such that a matrix of equal constraints is created, however, it is possible to tailor the precision requirements for each target variable and domain respectively. The population size of chromosomes to be generated is set to 20, the default mutation and elitist settings are used, the minimum number of units in each stratum is 2 and the number of iterations is set to 400. Again, these can vary according to the practical requirements of the experiments. With the common data frames, atomic strata, precision constraints as well as the above parameters the two algorithms are now ready to work. 

Each algorithm was evaluated 30 times with the default mutation and default elitist settings.  The results for Sample Size and Strata after 30 experiments each with 400 iterations are summarised in the scatterplot. 
%and numerical summary below.

\sisetup{
 round-mode          = places, % Rounds numbers
  round-precision     = 2, % to 2 places
}
\begin{figure}[H]
\centering{\label{fig:first}\includegraphics[]{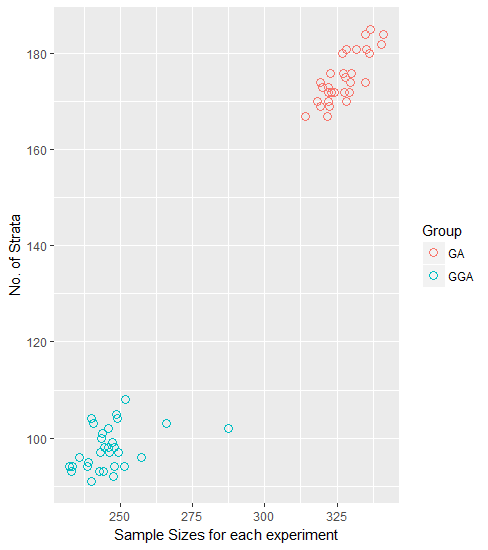}} 
    \caption{Scatterplot of Results for Strata v Sample size for GA and GGA after 30 experiments}
\end{figure}

% \usepackage{booktabs}
%\begin{table}[H]
%\centering
%\begin{tabular}{@{}lll@{}}
%\toprule
% & \textbf{GA} & \textbf{GGA} \\ \midrule
%\textbf{Min} & 314.1 & 232.5 \\
%\textbf{1st Quartile} & 321.9 & 240.2 \\
%\textbf{Median} & 327.3 & 245.2 \\
%\textbf{Mean} & 327.2 & 246.1 \\
%\textbf{3rd Quartile} & 331.3 & 248.5 \\
%\textbf{Max} & 341.2 & 287.5 \\
%\textbf{Interquartile range} & 9.37 & 8.24 \\ \bottomrule
%\end{tabular}
%\caption{Table 5.1 : Numerical summaries for GA and GGA}
%\end{table}

The scatterplot clearly shows that the GGA returns a smaller sample size to the GA for these settings. The median for the GGA, 246, is 25\% lower than that for the GA, 328.  We can conclude that for the above experiments the GGA yields lower sample size than the GA in 30 tests where the precision constraints were 0.05 for each target variable and for each domain, the default mutation and elitist settings are used, the population size of chromosomes was 20 and the number of iterations was 400.

\section{A comparison of the algorithms where the optimum solution is known}

There are differences between the mean sample size for each genetic algorithm. The next question is whether the new genetic algorithm can find the global minimum sample size.  Another consideration is whether the algorithm has been over-trained on the \emph{swissmunicpalities} data. Will it work on other data?  One way of dealing with such questions is to test it on another dataset example where the number of possible partitions is sufficiently small that the new algorithm can be tested to determine whether it finds the minimum sample size.  As the algorithm is searching for the minimum sample size through various recombinations of groupings of atomic strata, the more atomic strata there are the more the possible groupings there are.  Thus for a large number of atomic strata the number of possible partitions is so large that even with powerful computers it could be quite time consuming.

\citep{barcaroli2014samplingstrata} use the iris dataset\citep{anderson,fisher,R} to demonstrate that the GA they propose can find the optimum stratification i.e. the stratification or grouping of atomic strata which finds the minimum sample size.  The iris dataset is small and is widely available.  It has 150 observations for 5 variables Sepal Length, Sepal Width, Petal Length, Petal Width and Species.  Species is a categorical variable which has three levels, setosa, versicolor and virginica, each of which have 50 observations.

The remaining four variables are continuous measurements for length and width in centimetres. \citep{barcaroli2014samplingstrata} select Petal Length and Petal Width as variables of interest, i.e. target variables.  They select Sepal Length and Species as two auxiliary variables.  Sepal Length is continuous and, as is required for the algorithm, is converted to a categorical variable.  Therefore there are 8 usable atomic strata for this example.  The initial atomic strata are reproduced below:

\begin{table}[H]
\centering

\sisetup{
 round-mode          = places, % Rounds numbers
  round-precision     = 2, % to 2 places
}
%\begin{adjustbox}
\scriptsize
\begin{tabular}{@{}lllllllllll@{}}
%\toprule

\textit{$Stratum$} & \textit{$N$} & \textit{$M1$} & \textit{$M2$} & \textit{$S1$} & \textit{$S2$} & \textit{$X1$} & \textit{$X2$} & \textit{$DOMAIN$} \\
 &  &  &  &  &  &  &  &  \\
{[}4.3; 5.5{]} (1)*setosa & 45 & 1.466667 & 0.2444444 & 0.1712698 & 0.106574 & {[}4.3; 5.5{]} (1) & setosa & 1 \\
{[}4.3; 5.5{]} (1)*versicolor & 6 & 3.583333 & 1.1666667 & 0.4913134 & 0.2054805 & {[}4.3; 5.5{]} (1) & versicolor & 1 \\
{[}4.3; 5.5{]} (1)*virginica & 1 & 4.5 & 1.7 & 0 & 0 & {[}4.3; 5.5{]} (1) & virginica & 1 \\
{[}5.5; 6.5{]} (2)*setosa & 5 & 1.42 & 0.26 & 0.1720465 & 0.08 & {[}5.5; 6.5{]} (2) & setosa & 1 \\
{[}5.5; 6.5{]} (2)*versicolor & 35 & 4.268571 & 1.32 & 0.3670511 & 0.1894353 & {[}5.5; 6.5{]} (2) & versicolor & 1 \\
{[}5.5; 6.5{]} (2)*virginica & 23 & 5.230435 & 1.9478261 & 0.3181943 & 0.2887297 & {[}5.5; 6.5{]} (2) & virginica & 1 \\
{[}6.5; 7.9{]} (3)*versicolor & 9 & 4.677778 & 1.4555556 & 0.1930905 & 0.106574 & {[}6.5; 7.9{]} (3) & versicolor & 1 \\
{[}6.5; 7.9{]} (3)*virginica & 26 & 5.876923 & 2.1076923 & 0.4948253 & 0.2285794 & {[}6.5; 7.9{]} (3) & virginica & 1 \\
%\bottomrule

\end{tabular}
%\end{adjustbox}
\caption{Table 7.1 : Reproduction of table of atomic strata for estimating the minimum sample size for the target variables of  iris dataset as found in \citep{barcaroli2014samplingstrata}, P.379 }

\end{table}

where:

$M_g$ refer to the means for the corresponding $Y_g$ values in each atomic stratum $l_k$\\ $S_g$ refer to the population standard deviations for the corresponding $Y_g$ values in each atomic stratum $l_k$ \\
$N$ \textbf{\emph{in the table above}} refers to the number of records in each atomic stratum $l_k$ (not to be confused with $N$ the number of records in the entire population.\\

There are 4,140 possible partitions of the 8 atomic strata. Consequently, it is possible to test the sample size for the entire search space using the \emph{bethel.r} function.  This has already been done\citep{ballin2013joint} and the minimum sample size is known to be 11.

This test can be used to determine whether the new genetic algorithm correctly finds the minimum sample size without exploring
the entire search space. The fixed cost is set as equal to 0 and the average cost of observing a unit in stratum h is set equal to 1.  The cost of a solution, is therefore equal to the sample size.  The minimal sample size is the optimal solution.  The precision constraints for the two target variables are set to 0.05.

\begin{equation}
\begin{array}{rrclcl}

\displaystyle \min &\displaystyle C & = & n  = &  \sum_{h=1}^{H} n_h \\

\\  

\textrm{s.t.} & CV(T_g) & \leq & 0.05 \\

\end{array}
\end{equation}

The population size is set to 10 chromosomes.  The elitism rate for the generational algorithms is set to 0.2 and the mutation probability to 0.05.  For these tests the \emph{bethel.r} function will search for the minimum sample size, in integers rather than real numbers, for the grouping of atomic strata in each chromosome in the population.  The chromosomes will then be ranked by sample size in ascending order.

Accordingly the elite chromosomes are taken into the next iteration and the remaining chromosomes are generated using the recombination method for each algorithm.

The formula below calculates the number of new chromosomes generated over the course of each algorithm:

\begin{equation}
\begin{array}{rrclcl}

\\

N_{Chrom}= \left(\begin{array}{c}N_P\ + (N_P - E)\ . \ (N_{iters} -1)\end{array}\right) 

\\

\\

\end{array}
\end{equation}

where $N_{Chrom}$ is the number of chromosomes generated by generational algorithm, $N_{P}$ is the population size, $E$ is the number of elite chromosomes and $N_{iters}$ is the number of iterations.

\begin{figure}[H]
\centering{\label{fig:first}\includegraphics[]{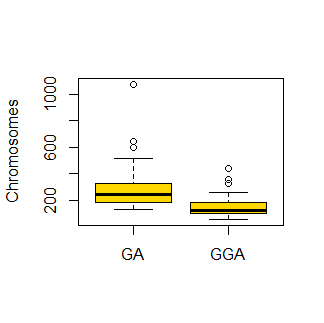}} 
    \caption{Boxplot of Chromomomes generated to find n=11 for GA and GGA after 30 experiments}
\end{figure}

%\end{adjustbox}

Figure 7.1 indicates that the grouping genetic algorithm finds the optimal solution more often than the genetic algorithm with fewer chromosomes generated.  These results also demonstrate that the new algorithm has been shown to work on a different dataset to the training data set and can find the optimum solution.

%As can seen from Table 7.2 the expected number of chromosomes required to find n=11 for the GA is 316 and for the GGA is 155 (rounding up).  This amounts to a 51\% difference in expected values.
%
%\begin{table}[H]
%\centering
%\begin{tabular}{lll}
%\hline
%                             & \textbf{GA} & \textbf{GGA} \\ \hline
%\textbf{Min}                 & 132         & 52           \\
%\textbf{1st Quartile}        & 188         & 100          \\
%\textbf{Median}              & 244         & 124          \\
%\textbf{Mean}                & 315.5       & 154.9        \\
%\textbf{3rd Quartile}        & 320         & 180          \\
%\textbf{Max}                 & 1076        & 436          \\
%\textbf{Interquartile range} & 132         & 80           \\ \hline
%\end{tabular}
%\caption{Table 6.2: Numerical summaries of Chromosomes generated to find n=11 for GA and GGA}
%\end{table}

The \emph{evalSolution.r} function in \emph{SamplingStrata} is used to take 50 samples from the population frame when the grouping or stratification that yields the optimal sample size is applied.  The average of the coefficients of variations provides the expected precision value which is compared with the accuracy constraints of 0.05 for each target variable that were set with objective function.

The results are displayed in Table 7.2, and suggest that both algorithms achieve precisions for the two target variables that are within the constraints of 0.05.

\begin{table}[H]

\centering
\tiny
\label{my-label}

\begin{tabular}{@{}lllll@{}}

\multicolumn{2}{l}{\textbf{(a) GA}} &           & \multicolumn{2}{l}{\textbf{(b) GGA}} \\ \midrule

\textbf{CV1}       & \textbf{CV2}       & \textbf{} & \textbf{CV1}        & \textbf{CV2}       \\ \midrule

0.027601           & 0.04967            &           & 0.027777            & 0.042311           \\ \bottomrule

\end{tabular}

\caption{Table 6.3 : Expected CVs for GA and GGA after 50 samples}

\end{table}

% \section{An application of the algorithms to the 2012 Commodity Flow Survey (CFS) Public Use Microdata (PUM) file }

\section{2015 American Community Survey Public Use Microdata}

The United States has been conducting a decennial census since 1790. In the 20th century censuses were split into long and short form versions. A subset of the population was required to answer the longer version of the census, with the remainder answering the shorter version. After the 2000 census the longer questionnaire became the annual American Community Survey(ACS)\citep{USCB:2013}. The 2015 ACS Public Use Microdata Sample (PUMS) file is a sample of actual responses to the American Community Survey (ACS) representing 1\% of the US population. The dataset contains variables which represent nearly every question in the survey. There are also new variables which are derived from multiple survey responses after the survey. The PUMS file contains 1,496,678 records each of which represents a unique housing unit or group quarters. There are 235 variables. The full data dictionary is available in \cite{USCB:2016}. The PUMS data contains records for which there are missing values for some variables. We would like to start with a population complete sampling frame and our approach to deal with missing values is as follows.

We selected the following to be target variables. 

\begin{itemize}
\item 1. Household income (past 12 months)
\item 2. Property value
\item 3. Selected monthly owner costs
\item 4. Fire/hazard/flood insurance (yearly amount).
\end{itemize}

The following auxiliary variables were selected :
\begin{itemize}
\item 1. Units in structure
\item 2. Tenure
\item 3. Work experience of householder and spouse
\item 4. Work status of householder or spouse in family households
\item 5. House heating fuel
\item 6. When structure first built. 
\end{itemize}
The auxiliary variables were categorical, so no clustering technique was needed.

Once the target and auxiliary variables were selected, a subset of the PUMS data for which all values were present for all selected variables were present was then read into R. This subset had 619,747 records. %There are possible imputation methods such as filling missing values with the median or winsorized mean, using prior survey data, hot-deck imputation, bootstrapping, or using principal component analysis where the scores are multiplied by the loadings to reconstruct the data matrix. We felt that use of imputation techniques to create a fully populated sampling frame required further work.  

In keeping with the above analysis on the \emph{swissmunicipalities} data the upper limits for the Coefficients of Variation for the target variables were set to 0.05.  Similarly, 30 experiments were conducted for each algorithm. The cost of a solution is equal to the sample size. The minimal sample size is the optimal solution.  The population size was set to 20, the number of iterations was set to 400, elitism was set to be 20\%, and the default mutation probability was used.

Table 7.2 shows the mean sample size and strata for each of the algorithms. It is clear that the generational grouping genetic algorithm has produced significantly lower sample sizes to the genetic algorithm.  The mean sample size given by the grouping genetic algorithm is 17,547 which is 80\% lower than that for the genetic algorithm, 91,121. The mean number of strata, 1,830, which was found by the GGA, is 96\% lower than the mean 49,722 found by the GA. 

For illustration purposes only, estimated savings are tabulated using the most recent available average sampling costs from the 2013 American Community Survey \citep{ESA:2015}.  
The assumption is that the average sampling costs would be the same for the survey specifications tested in the above experiments. In Table 7.3 a saving of \$5 million could be attained.

\begin{table}[H]
\tiny
\centering
\begin{tabular}{lllll}
\cline{1-2} \cline{4-5}
\multicolumn{2}{l}{\textbf{(a) GA}} &  & \textbf{(b) GGA} & \textbf{} \\ \cline{1-2} \cline{4-5} 
\textbf{Sample Size} & \textbf{Strata} &  & \textbf{Sample Size} & \textbf{Strata} \\ \cline{1-2} \cline{4-5} 
91121 & 49722 &  & 17547 & 1830 \\ \cline{1-2} \cline{4-5} 
\end{tabular}
\caption{Table 7.1 : Mean sample size and strata for the GA and the GGA}
\end{table}

\begin{table}[H]
\tiny
\centering

\label{my-label}

\begin{tabular}{@{}llllll@{}}
\hline
\textbf{ACS Survey Year} & \textbf{Budget  (\$ million)} & \textbf{Sample Size} & \textbf{Average Cost per Sampling Unit  (\$ million)} & \textbf{Difference in Sample Sizes} & \textbf{Savings (\$ million)} \\ \hline
2013 & 242 & 3,540,000 & 0.00006836 & 73,574 & 5.029574 \\ \hline
\end{tabular}
\caption{Table 7.2 : Budget and sample size for the 2013 edition of the ACS}
\label{my-label}
\end{table}

%\paragraph*{Performance of the Genetic Algorithms}\\
\subsection*{Performance of the Genetic Algorithms}
 
The convergence plots below show the significant impact the
default mutation probability for these experiments has on the
performance of the algorithms.  The convergence plots are for the
output for the first experiment ran for each algorithm.  The black
line represents the best or lowest sample size for the population in
each iteration, whereas the red line represents the mean sample size
for the population in each iteration.

\begin{figure}[H]
  \centering
\subfloat[GA]{\includegraphics[width=2.5in]{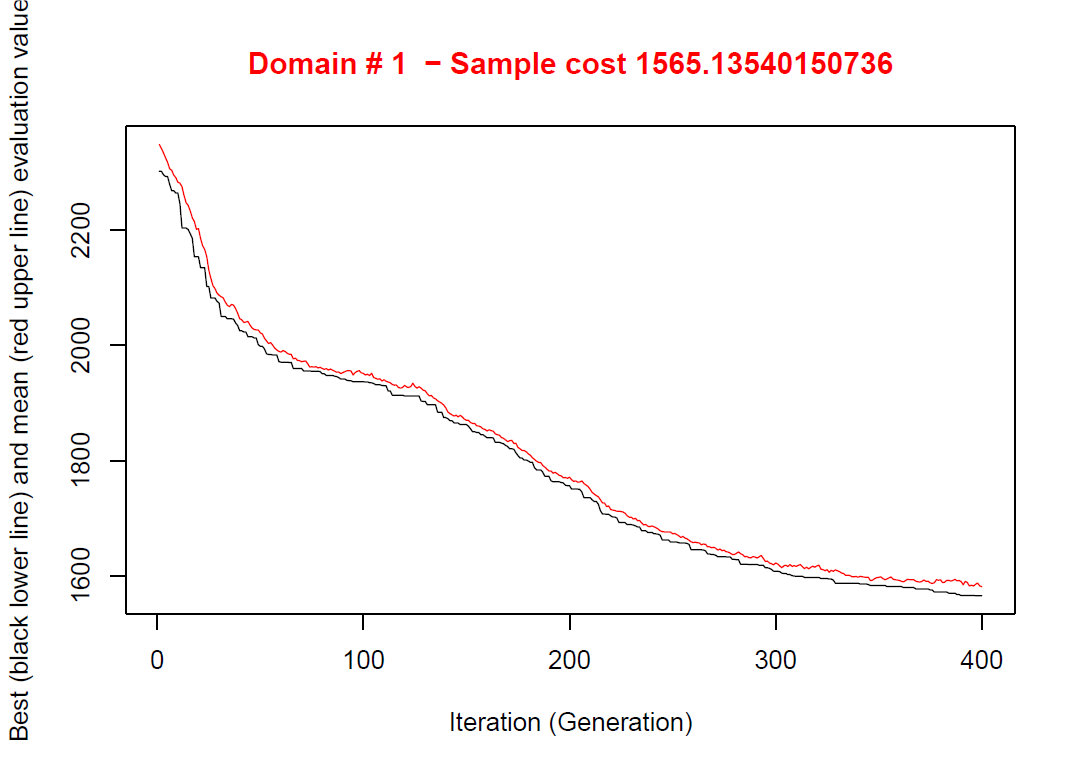}}
  \hfill
\subfloat[GGA]{\includegraphics[width=2.5in]{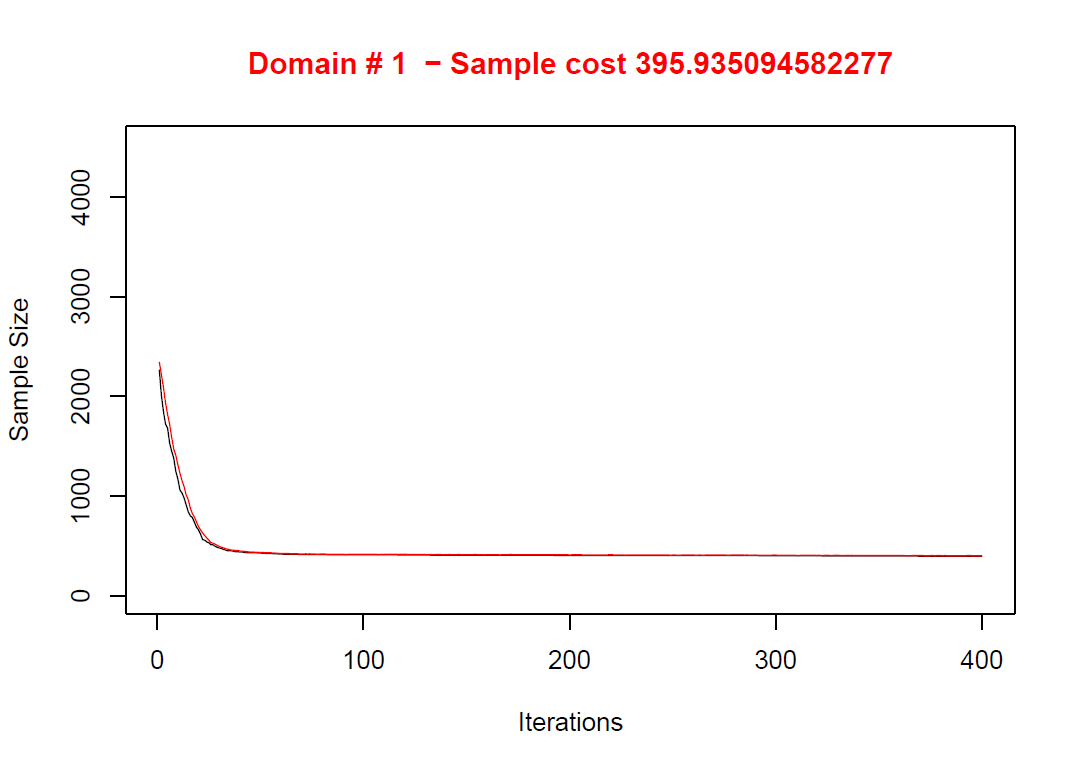}}
\caption{ Convergence plots for Sample Size after the 1st experiment for GA and GGA }
\end{figure}

The GA appears to be performing steadily with the default mutation
probability and does not appear to have reached a local minimum.  The GGA appears to have reached a local or global minimum very quickly. The results of the algorithms might be improved by optimising the mutation settings.  Nonetheless for the small population size of 20 and after only 400 iterations it has shown significant savings on sample size when compared to the \emph{SamplingStrata} genetic algorithm.

\section{Further Tests}
We have thus far demonstrated that both algorithms can find the optimum solution and that as datasets get bigger and the Cartesian products of the auxiliary variables get more complex the grouping genetic algorithm converges on a local minimum faster than the original genetic algorithm. Indeed as can be seen from Figure 8.1 the GGA appears to have converged on a solution in less than 100 iterations, whereas the GA appears to be still in descent. We will now compare the performance of the GGA with the GA on alternative sampling frames using 100 iterations, with the other starting parameters remaining the same. We hope the results will further corroborate our claim that our GGA can more efficiently partition the atomic strata than the GA.  

\subsection{Kaggle Data Science for Good challenge Kiva Loans data}

The online crowdfunding platform kiva.org provided a dataset of loans issued to people living in poor and financially excluded circumstances around the world over a two period for a Kaggle  Data Science for Good challenge. The dataset has 671,205 unique records.
\\
\\
We selected these target variables: 

\begin{itemize}[label={}]
\item 1.	term in months
\item 2. 	lender count
\item 3.	loan amount
\end{itemize}

We then used the following auxiliary variables: 

\begin{itemize}[label={}]
\item 1.	sector
\item 2. 	currency
\item 3.	activity
\item 4.    region
\item 5.    partner id
\end{itemize}

to create atomic strata. The domain for this experiment was country code. For these variables we removed any records with missing values. We then proceeded to remove any countries with less than 10 records from the sampling frame. This resulted in a sampling frame with 614,361 records. We again used 100 iterations, along with the default starting parameters, i.e. Coefficients of Variation of 0.05, the cost of a solution is equal to the sample size, the population size was set as 20, elitism was 20\% and we again used the default mutation probability. The variable country\_code was selected as domain of which there were 73 unique entries. The algorithms were searching for the smallest sample size for each of these.  

\begin{table}[H]
\centering
\begin{tabular}{llllll}
\multicolumn{2}{l}{\textbf{GA}} & \multicolumn{2}{l}{\textbf{GGA}} & \multicolumn{2}{l}{\textbf{Reduction}} \\
\textbf{Sample size} & \textbf{Strata} & \textbf{Sample size} & \textbf{Strata} & \textbf{Sample size} & \textbf{strata} \\
78018 & 43030 & 11963 & 1793 & 84.67\% & 95.83\%
\end{tabular}
\end{table}

The above table shows an 84.67\% in sample size and a 95.83\% reduction in the number of strata after 100 iterations. 

%https://www.kaggle.com/unitednations/global-commodity-trade-statistics
\subsection{UN Commodity Trade Statistics data}

Kaggle also hosts a copy of the UN Statistical Division Commodity Trade Statistics data. Trade records are available from 1962. We took a subset of data for the year 2011 and removed records with missing observations. This resulted in a data set with 351,057 records

We selected the following target variable

\begin{itemize}[label={}]
\item 1.	trade\_usd
\end{itemize}

The variable trade\_usd refers to the value of trade in USD. 
We then selected the following auxiliary variables: 

\begin{itemize}[label={}]
\item 1.	commodity
\item 2. 	flow
\item 3.	category
\end{itemize}

The variable commodity is a categorical description of the type of commodity, e.g. Horses, live except pure-bred breeding. The variable flow describes whether the commodity was an import, export, re-import or re-export. The variable category describes the category of commodity, e.g. silk or fertilisers. The 171 categories of country or area were selected as domains. The same parameters were used. 

\begin{table}[H]
\centering
\begin{tabular}{llllll}
\multicolumn{2}{l}{\textbf{GA}} & \multicolumn{2}{l}{\textbf{GGA}} & \multicolumn{2}{l}{\textbf{Reduction}} \\
\textbf{Sample size} & \textbf{Strata} & \textbf{Sample size} & \textbf{Strata} & \textbf{Sample size} & \textbf{strata} \\
288638 & 191000 & 84181 & 16555 & 70.84\% & 91.33\%
\end{tabular}
\end{table}

This data example was a subset of the commodity trade statistics data for 2011. It had less records than ACS or Kiva loans data. We need to be mindful of the smaller number of records and the greater number of domains. However, there was a 70.84\% reduction in sample size and a 91.33\% reduction in strata after 100 iterations.

\subsection{2000 US census data}

The Integrated Public Use Microdata Series extract is a 5\% sample of the 2000 US census data \cite{ruggles2015integrated}. The file contains 6,184,483 records. The US Census Data will be very similar to the American Community Survey data as the latter is an annual version of the former. As census data tends to be used as sampling frames for household surveys, we thought it would be interesting to compare the algorithms on the US census data for a different year to the ACS and for different target and auxiliary variable combinations. 

The one target variable in this test is usually a key focus of household surveys.
\begin{itemize}[label={}]
\item 1.	Total household income
\end{itemize}

We used the following information as auxiliary variables (note these are variables which are likely available in administrative data):
\begin{itemize}[label={}]
\item 1.	Annual property insurance cost
\item 2. 	Annual home heating fuel cost
\item 3.	Annual electricity cost
\item 4.    House value
\end{itemize}

The house value variable (VALUEH) reports the midpoint of house value intervals (e.g. 5,000 is the midpoint of the interval of less than 10,000), so we have treated it as a categorical variable. As with the 2015 ACS Public Use Microdata Sample (PUMS) dataset we have taken a subset for which all values are present. For example we have removed all records where values for total household income are not applicable. The concept was to create a dataset where there was no missing values which could be used as a sampling frame to compare both algorithms. This has resulted in a subset with 627,611 records. The domain for this experiment was Census region and division. As before we have set the starting parameters as follows the upper limit for the Coefficients of Variation is 0.05, the cost of a solution is equal to the sample size, the population size for the algorithms is 20, elitism is retained at 20\% and we again used the default mutation probability. The results in table below show the difference in sample size and the number of strata after 100 iterations.

\begin{table}[H]
\centering
%\begin{adjustbox}{\textwidth}
\tiny\begin{tabular}{|l|l|l|l|l|l|l|l|l|}
\hline
\textbf{} & \multicolumn{2}{c|}{\textbf{Sampling frame}} & \multicolumn{2}{c|}{\textbf{GA solution}} & \multicolumn{2}{c|}{\textbf{GGA solution}} & \multicolumn{2}{c|}{\textbf{\% Difference in solutions}} \\ \hline
\multicolumn{1}{|c|}{\textbf{Division}} & \textbf{Sampling Units} & \textbf{Atomic Strata} & \textbf{Sample sizes} & \textbf{Strata} & \textbf{Sample sizes} & \textbf{Strata} & \textbf{Sample sizes} & \textbf{Strata} \\ \hline
\textbf{New England} & 116,045 & 87,084 & 81,012 & 52,628 & 375.84 & 58 & \textit{99.54\%} & \textit{99.89\%} \\ \hline
\textbf{Middle Atlantic} & 183,543 & 138,470 & 130,862 & 86,002 & 415.69 & 75 & \textit{99.68\%} & \textit{99.91\%} \\ \hline
\textbf{East North Central} & 65,480 & 58,055 & 53,075 & 35,794 & 326.95 & 42 & \textit{99.38\%} & \textit{99.88\%} \\ \hline
\textbf{West North Central} & 31,408 & 29,413 & 26,525 & 18,248 & 323.60 & 38 & \textit{98.78\%} & \textit{99.79\%} \\ \hline
\textbf{South Atlantic} & 97,189 & 83,357 & 76,716 & 51,457 & 439.29 & 49 & \textit{99.43\%} & \textit{99.90\%} \\ \hline
\textbf{East South Central} & 21,631 & 20,429 & 18,256 & 12,500 & 450.06 & 62 & \textit{97.53\%} & \textit{99.50\%} \\ \hline
\textbf{West South Central} & 22,582 & 20,919 & 18,750 & 12,730 & 406.92 & 39 & \textit{97.83\%} & \textit{99.69\%} \\ \hline
\textbf{Mountain} & 26,765 & 25,041 & 22,161 & 14,791 & 350.33 & 30 & \textit{98.42\%} & \textit{99.80\%} \\ \hline
\textbf{Pacific} & 62,968 & 54,864 & 50,136 & 33,653 & 357.21 & 49 & \textit{99.29\%} & \textit{99.85\%} \\ \hline
\textit{\textbf{Total}} & \textit{\textbf{627,611}} & \textit{\textbf{517,632}} & \textit{\textbf{477,493}} & \textit{\textbf{317,803}} & \textit{\textbf{3,445.88}} & \textit{\textbf{442}} & \textit{\textbf{99.28\%}} & \textit{\textbf{99.86\%}} \\ \hline
\end{tabular}
%\end{adjustbox}
\end{table}

The results show a sample size of 3,446 for the GGA and a sample size of 477,493 for the GA after 100 iterations. This represents a 99.28\% difference in sample size due to the efficiency gained by retaining good groupings of atomic strata employed by the GGA over the traditional method of random splitting of good groups used by the GA. %Notice the high correlation of the number of atomic strata with the number of sampling units in each division. We will discuss this further in section 9 below.
\section{Processing times}

Our GGA was proposed and developed so that it would work with the rest of the functions in \emph{SamplingStrata}.  Therefore the rest of the functions in the package remained unchanged.  This includes the function the \emph{bethel.r} function which evaluates the fitness of chromosomes in every iteration and is computationally expensive. Because the same function was evaluating chromosomes for each algorithm, the processing times were largely the same for all experiments on the above three datasets.  The experiments were run on three different computers, one with Windows 10 and the other two with Ubuntu 14.04.  Two computers were laptops and one was a desktop PC.  All computers had 8GB RAM.  For the 2015 American Flow Survey (ACS) Public Use Microdata (PUM) dataset, the experiments took approximately 30 days per genetic algorithm. Any improvement in processing time will mean that the increased number of iterations that can be run for the same time cost, could facilitate further searching of the universe of solutions and could provide further cost savings. 

After we developed the GGA we then started work on making the evaluation algorithm faster. We searched for bottlenecks in bethel.r using the R \emph{lineprof} package. Our analysis of results suggested that the function within
bethel.r called chromy appears to take the bulk of computational time. A further examination reveals that chromy contains a while loop with a default setting of 200 iterations. Furthermore bethel.r itself can be run on each candidate solution in any population on a dataset of any functional size (which we have the computation power to process) for any number of iterations. Bigger datasets will take longer to process. We felt performance could made faster by converting bethel R algorithm into to C++ and then intergrating the C++ version into R using the \emph{Rcpp} package. 

\begin{table}[H]
\centering
\tiny
\begin{tabular}{lllllll}
\textbf{Dataset} & \textbf{Records} & \textbf{Domains} & \textbf{\begin{tabular}[c]{@{}l@{}}Atomic \\ Strata\end{tabular}} & \textbf{\begin{tabular}[c]{@{}l@{}}Bethel/\\ microseconds\end{tabular}} & \textbf{\begin{tabular}[c]{@{}l@{}}BethelRcpp2018/\\ microseconds\end{tabular}} & \textbf{\begin{tabular}[c]{@{}l@{}}Speed-up \\ Factor\end{tabular}} \\
\textit{\textbf{iris}} & 150 & 1 & 8 & 2684.7715 & 143.1335 & 18.76 \\
\textit{\textbf{swissmunicipalities}} & 2896 & 7 & 641 & 99916.17 & 10749.51 & 9.29 \\
\textit{\textbf{American Community Survey 2015}} & 619747 & 51 & 123007 & 565278500 & 47858200 & 11.81 \\
\textit{\textbf{US Census Data 2000}} & 627611 & 9 & 517632 & 2686771 & 1303667 & 2.06 \\
\textit{\textbf{Kiva Loans Data}} & 614361 & 73 & 84897 & 826297710 & 82894480 & 9.97 \\
\textit{\textbf{UN Commodity Trade Data 2011}} & 351057 & 171 & 350895 & 139749810 & 87555870 & 1.60
\end{tabular}
\end{table}

The table shows the median time taken to run the bethel algorithm one hundred times for the datasets we used to conduct our analysis. Our results show that the C++ version of bethel is faster than the R version. The speed up could make a practical difference in the number of iterations that can be run in \emph{SamplingStrata} due to the processing times required for \emph{bethel.r}. However, performance will vary according to the size and complexity of the problem (assuming the computer is not devoting memory to other demanding tasks at the same time). The speed up is achieved because C++ communicates at a lower level with the computer than R. However, it is also due to the complexity of the analysis conducted in each for loop as well as the fact that larger data will restrict the available memory .

\section{Conclusions}

We created a GGA to be an alternative to the \emph{SamplingStrata} GA. We then compared the two algorithms using a number of datasets. Statistical tests show that the mean results for 30 tests of the new algorithm are different to 30 tests for the \emph{SamplingStrata} genetic algorithm for the \emph{swissmunicpalities} data.

Furthermore, in the \emph{iris} dataset example they were shown to find the correct results where the solution is known.  When an average of 50 samples were selected from the groupings of atomic strata proposed by the GGA for the \emph{iris} data test, the accuracy was on average below the upper constraint.  Clear cost savings were achieved with the generational grouping genetic algorithms when compared with the genetic algorithms for the same number of iterations for the 2015 American Community Survey (ACS) Public Use Microdata (PUM) dataset. Additional tests showed a marked reduction in sample size for the Kaggle Kiva Loans data, Kaggle 2011 UN Commmodity Trade data and the US Census 2000 data.

\section{Further work}
The GGA compares favourably with the GA at finding the correct solution and meeting constraints, but outperforms the GA on larger datasets where the number of iterations has been restricted.  This is useful for datasets where the number of iterations has to be constrained owing to computational burden. We have reported faster processing times by integrating the behtel.r function with C++ using the Rcpp package. However, alternative evaluation techniques that will speed up the runtimes of the algorithm could be considered. Further research could also be undertaken into other machine learning techniques for solving this problem. Finally, imputation techniques to fill in missing data and thus create a complete sampling frame could be considered.

\section*{Acknowledgement}

We wish to acknowledge Steven Riesz of the Economic Statistical Methods Division of the U.S Census Bureau and Brian J Mc Elroy of the Economic Reimbursable Survey Division of the U.S Census Bureau both of whom answered questions which were of assistance in choosing which US Census Bureau data to use. We would also like to thank Giulio Barcaroli and Marco Ballin co-authors of \citep{ballin2013joint} for independently testing our GGA. 
%----------------------------------------------------------------------------------------
%	BIBLIOGRAPHY
%----------------------------------------------------------------------------------------

%\section*{References}
\bibliographystyle{chicago}
%\RaggedRight
\Urlmuskip=0mu plus 1mu\relax
\def\UrlBreaks{\do\/\do-}
%\bibliography{arxiv3_2018} % The file containing the bibliography

%----------------------------------------------------------------------------------------

%\section*{Appendix}
%
%\begin{lstlisting}[language=R]
%library(SamplingStrata)
%frame<-NULL
%frame$Y1<-c(57324,32429,28161,31223,24321,45621)
%frame$Y2<-c(1314,6007,5034,2344,1632,6756)
%frame$X1 <-c(18,17,17,19,19,19)
%frame$X2 <- c(3,1,1,1,1,1)
%frame$domainvalue<-c(2,1,1,2,2,2)
%frame <- as.data.frame(frame)
%strata <- buildStrataDF(frame)
%strata
%\end{lstlisting}
\end{document}